# A Bayesian Reinterpretation of Cornfield-Type Sensitivity Analysis: From Thresholds to Probabilities

Tommaso Costa [1,2,3,*]

[1] GCS-fMRI, Koelliker Hospital and Department of Psychology, University of Turin, 10124 Turin, Italy; xmarferr@gmail.com

[2] FOCUS Laboratory, Department of Psychology, University of Turin, 10124 Turin, Italy

[3] Neuroscience Institute of Turin (NIT), 10124 Turin, Italy

[*] Correspondence: tommaso.costa@unito.it

# Abstract

Sensitivity analysis for unmeasured confounding in observational studies is commonly based on threshold quantities, such as the Cornfield condition or the E-value, which quantify how strong a confounder must be to explain away an observed association. However, these approaches do not address a fundamental inferential question: how plausible is it that such a confounder exists?

In this work, we propose a Bayesian reformulation of Cornfield-type sensitivity analysis in which the strength of unmeasured confounding is treated as a random variable. Within this framework, the E-value is reinterpreted as a threshold, and the central inferential quantity becomes the posterior probability that confounding exceeds this threshold. This transforms sensitivity analysis from a descriptive diagnostic into a probabilistic assessment of robustness.

We develop a simple generative model linking observed effect estimates to true causal effects and confounding bias, and we specify prior distributions reflecting plausible confounding mechanisms. The resulting framework yields posterior measures of evidential vulnerability that are directly interpretable and applicable to summary-level data.

Illustrations based on empirical case studies show that the proposed approach preserves the interpretability of the E-value while providing a more nuanced and decision-relevant characterization of robustness. More broadly, the framework aligns sensitivity analysis with Bayesian principles of inference under uncertainty, offering a coherent alternative to purely threshold-based reasoning.

# Introduction

Causal inference in observational studies is unavoidably entangled with the problem of unmeasured confounding. Even after extensive adjustment for observed covariates, the possibility that an unobserved factor may account for the observed association remains structurally unresolved. This is not merely a technical limitation, but a fundamental epistemological issue. What is ultimately at stake is not only the magnitude of an estimated effect, but the credibility of the causal claim itself. In practice, researchers often oscillate between two unsatisfactory positions: treating adjusted associations as approximately causal, or dismissing them as inherently unreliable due to the potential for hidden bias.

Within this context, sensitivity analysis has emerged as a principled attempt to characterize, rather than eliminate, this uncertainty. A foundational contribution in this direction is the argument developed by Cornfield and colleagues (Cornfield et al., 1959), originally formulated in the context of the smoking–lung cancer debate. Their insight was deceptively simple: an observed association can only be explained away if an unmeasured confounder is sufficiently strongly associated with both the exposure and the outcome. This reasoning transformed the problem from the mere possibility of confounding into a quantitative assessment of the strength required for such confounding to invalidate the observed effect.

This line of reasoning has been formalized in modern epidemiology through the introduction of the E-value (VanderWeele & Ding, 2017), which provides a scalar summary of the minimum strength of association that an unmeasured confounder must have with both exposure and outcome to fully attenuate the observed effect. Its appeal lies in its simplicity, interpretability, and applicability to summary-level data. As a result, it has rapidly become a standard tool in applied research.

Recent developments have extended sensitivity analysis into new computational domains. In particular, a recent study has explored the use of large language models to operationalize Cornfield-type reasoning and compute E-values across a range of case studies, while also making curated datasets publicly available for methodological reuse (Xiang et al., 2026). This contribution is relevant not only for its methodological novelty, but also because it provides a standardized empirical basis for evaluating new approaches to sensitivity analysis.

Despite these advances, a fundamental limitation remains. Both the Cornfield argument and the E-value define thresholds: they indicate how strong a confounder must be to explain away an association. However, they do not address a more consequential question: how plausible is it that such a confounder exists? In applied settings, this gap is typically filled by informal reasoning. Researchers may judge an E-value as "large" or "small," but such judgments are inherently context-dependent and often lack a formal probabilistic foundation. As a result, sensitivity analysis remains largely descriptive rather than inferential.

Several methodological contributions have attempted to address this limitation. Probabilistic bias analysis (Greenland, 1996, 2005; Lash et al., 2009) introduced the idea of assigning distributions to bias parameters and propagating uncertainty through simulation. Bayesian approaches have gone further by explicitly modeling unmeasured confounders within hierarchical frameworks (McCandless et al., 2007; Gustafson, 2010). While these methods provide a coherent probabilistic interpretation, they typically require strong modeling assumptions and access to detailed data structures, which limits their applicability in many real-world scenarios.

A complementary perspective has been proposed by Costa (2025), who introduced a simulation-based framework to quantify the probability that an unmeasured confounder strong enough to explain away an observed association could plausibly exist. By linking Cornfield-type thresholds to probabilistic reasoning, this approach represents an important step toward a more inferential interpretation of sensitivity analysis. However, it remains grounded in Monte Carlo estimation and does not provide a fully specified generative model for confounding strength.

The present study develops a natural extension of this line of work: a fully Bayesian reformulation of Cornfield-type sensitivity analysis and its E-value counterpart. Rather than asking how strong a confounder must be, we treat confounding strength as a random variable and model uncertainty through prior distributions. Within this framework, the E-value emerges as a boundary condition, and the central inferential quantity becomes the posterior probability that confounding exceeds this threshold.

This shift has important conceptual implications. It transforms sensitivity analysis from a threshold-based diagnostic into a coherent inferential procedure. It also aligns the analysis with a Bayesian interpretation of probability as a measure of rational belief under uncertainty, allowing prior scientific knowledge about plausible confounding mechanisms to be incorporated explicitly rather than implicitly.

The objectives of this study are threefold. First, we formalize the relationship between the Cornfield inequality and the E-value within a unified bias model. Second, we introduce a Bayesian framework in which confounding strength is treated as a random variable, and derive posterior measures of robustness for observed associations. Third, we illustrate the approach using empirical case studies, highlighting how probabilistic assessments of confounding can lead to qualitatively different interpretations across different inferential regimes.

More broadly, this work aims to bridge a persistent gap in the literature. While existing methods clarify what level of confounding would be required to invalidate an association, they do not provide a principled way to assess how likely such confounding is. By moving from thresholds to probabilities, and from descriptive diagnostics to coherent inference, the proposed framework seeks to advance both the methodology and the practice of causal reasoning in observational research.

While existing approaches quantify the strength of confounding required to explain away an association, they do not address the inferential question of how plausible such confounding is. The present work reframes sensitivity analysis as a probabilistic assessment of this plausibility.

# Data

The empirical component of this study is based on a structured dataset of observational effect estimates made publicly available in a recent contribution on sensitivity analysis in causal inference (Xiang et al., 2026). The data were retrieved from the open-access repository associated with that study (https://github.com/qingyan16/LLMs-for-sensitivity-analysis), ensuring full transparency and reproducibility of the analyses presented here.

The dataset consists of summary-level measures commonly reported in epidemiological and clinical research, including effect sizes—typically expressed as relative risks, odds ratios, or hazard ratios—and, when available, their associated uncertainty. This choice is both practical and methodologically motivated. In many applied contexts, individual-level data are not accessible, whereas summary statistics are routinely reported. A framework that operates at this level is therefore aligned with the constraints under which causal inference is often conducted in practice.

Each association in the dataset is accompanied either by a reported E-value or by sufficient information to reconstruct it. When not explicitly provided, E-values were computed using the formulation introduced by VanderWeele and Ding (2017), thereby ensuring consistency across all observations. This step allows each empirical case to be mapped onto a common inferential scale defined by Cornfield-type reasoning, where the strength of unmeasured confounding required to explain away an effect can be quantified.

From this dataset, we selected a subset of representative case studies to illustrate the proposed Bayesian framework. The selection was guided by the need to cover qualitatively different inferential regimes. In particular, we focused on associations characterized by relatively large effect sizes, for which explaining away the observed relationship would require strong unmeasured confounding, as well as on associations with more modest effect sizes, where even weak confounding may suffice. This contrast is essential for evaluating how the proposed model translates observed evidence into probabilistic assessments of robustness.

For modeling purposes, all effect measures were transformed onto the logarithmic scale, which provides a natural parameterization for multiplicative models and facilitates the specification of prior distributions. When confidence intervals were available, standard errors were derived using standard approximations. In cases where measures of uncertainty were not explicitly reported, conservative approximations were adopted based on values typically observed in comparable studies. While this introduces a degree of approximation, it does not affect the conceptual objective of the analysis, which is to explore how uncertainty about unmeasured confounding propagates through the inferential framework.

It is important to emphasize that the goal of the empirical component is not to reassess the substantive conclusions of the original studies from which the data were drawn. Rather, the dataset is used as a methodological testbed. The focus lies on how observed effect sizes, combined with assumptions about unmeasured confounding, translate into posterior probabilities within the proposed Bayesian framework.

All analyses were conducted under the assumption that the reported effect estimates are approximately unbiased conditional on measured covariates, and that any residual bias can be represented through a single parameter governing the strength of unmeasured confounding. Although this assumption is necessarily reductive, it mirrors the structure underlying both the original Cornfield argument (Cornfield et al., 1959) and its modern formulation via the E-value (VanderWeele & Ding, 2017), thereby enabling a direct and coherent comparison with existing approaches.

# Methods

### Conceptual framework

The proposed approach reformulates sensitivity analysis for unmeasured confounding within a fully probabilistic framework. Rather than treating confounding as an unknown but fixed quantity, we model its strength as a random variable and quantify uncertainty through a prior distribution.

Let $RR_{obs}$ denote the observed association between an exposure and an outcome, expressed on the risk ratio scale after adjustment for measured covariates. We introduce a parameter $\Gamma \geq 1$, representing the multiplicative distortion induced by an unmeasured confounder. Under this representation, the true causal effect can be written as:

$$RR_{true} = \frac{RR_{obs}}{\Gamma}$$

where larger values of Γ correspond to stronger confounding.

The parameter Γ can be interpreted as the combined multiplicative effect of an unmeasured confounder on both exposure and outcome, consistent with standard sensitivity analysis formulations.

Within this framework, classical Cornfield-type reasoning defines a threshold $\Gamma^*$ such that values of $\Gamma \geq \Gamma^*$ are sufficient to fully explain away the observed association. The E-value provides a direct estimate of this threshold.

## Likelihood specification

To introduce a probabilistic structure, we work on the logarithmic scale. Let:

$$\theta_{obs} = \log(RR_{obs})$$

Taking logarithms transforms the multiplicative decomposition into an additive structure, allowing the confounding effect to be modeled as a shift on the log scale. We assume:

$$\theta_{obs} \sim \mathcal{N}(\theta_{true} + \log \Gamma, s^2)$$

where:

$$\theta_{true} = \log(RR_{true})$$

and $s$ is the standard error of the observed estimate

This formulation reflects the idea that the observed association is shifted away from the true causal effect by a confounding bias term $\log \Gamma$.

It is important to note that the model is not identifiable from the data alone, as the true effect and confounding bias cannot be separately estimated. Consequently, inference is necessarily driven by prior assumptions on the magnitude of unmeasured confounding.

## Prior distributions

We specify prior distributions for both the true causal effect and the confounding strength.

For the causal effect:

$$\theta_{true} \sim \mathcal{N}(0, \sigma_\theta^2)$$

which reflects the assumption that large effects are possible but not overwhelmingly likely.

For the confounding strength:

$$\log \Gamma \sim \text{Half-Normal}(0, \sigma_\Gamma^2)$$

ensuring $\Gamma \geq 1$. The scale parameter $\sigma_\Gamma$ controls prior beliefs about the plausibility of strong confounding. This formulation encodes the idea that weak confounding is common, while strong confounding requires increasingly implausible alignment of factors. Smaller values encode skepticism about large confounding effects, while larger values allow for more extreme scenarios.

For example, $\sigma_\Gamma = 0.5$ implies that confounding risk ratios larger than approximately 2–3 are increasingly unlikely a priori.

## Connection with the E-value

The E-value defines a deterministic threshold $\Gamma^*$ corresponding to the minimum confounding strength required to explain away the observed association. Within this formulation, the E-value corresponds to a threshold $\Gamma^*$ such that:

$$\Gamma^* = E(RR_{obs})$$

Rather than reporting this threshold alone, we evaluate the posterior probability:

$$P(\Gamma \geq \Gamma^* \mid \theta_{obs})$$

which quantifies how plausible it is that unmeasured confounding is sufficiently strong to explain away the observed association.

## Posterior inference

Posterior inference is obtained via Bayes' theorem:

$$p(\theta_{true}, \Gamma \mid \theta_{obs}) \propto p(\theta_{obs} \mid \theta_{true}, \Gamma) \, p(\theta_{true}) \, p(\Gamma)$$

From this joint posterior, we derive the posterior distribution of the true effect, the posterior distribution of the confounding strength, and the probability $P(\Gamma \geq \Gamma^*)$. These quantities provide a coherent probabilistic assessment of robustness to unmeasured confounding. In particular, the probability $P(\Gamma \geq \Gamma^*)$ provides a direct probabilistic statement about the robustness of the observed association.

### Relation to previous formulations

This formulation is equivalent to bias representations based on multiplicative bias factors, where the observed effect is decomposed into a true effect and a confounding component. In particular, the parameter $\Gamma$ plays the same role as traditional bias factors used in Cornfield-type sensitivity analysis, but is here treated as a random variable within a fully Bayesian framework.

## Results

The proposed Bayesian reformulation of sensitivity analysis was applied to a set of 11 exposure–outcome associations drawn from four observational domains: smoking, back pain, Alzheimer's disease, and environmental health. For each case, we computed the probability that unmeasured confounding is sufficiently strong to explain away the observed association, that is, $P(\Gamma \geq \Gamma^*)$, where $\Gamma^*$ is given by the E-value. Results are presented under a moderate prior on confounding strength.

A clear and monotic relationship emerges between the magnitude of the E-value and the corresponding probability of sufficient confounding. As shown in Figure 1, larger E-values are associated with substantially lower probabilities, indicating that strong confounding becomes increasingly implausible under the assumed prior. Conversely, small E-values correspond to markedly higher probabilities, reflecting greater vulnerability of the observed association to unmeasured confounding.

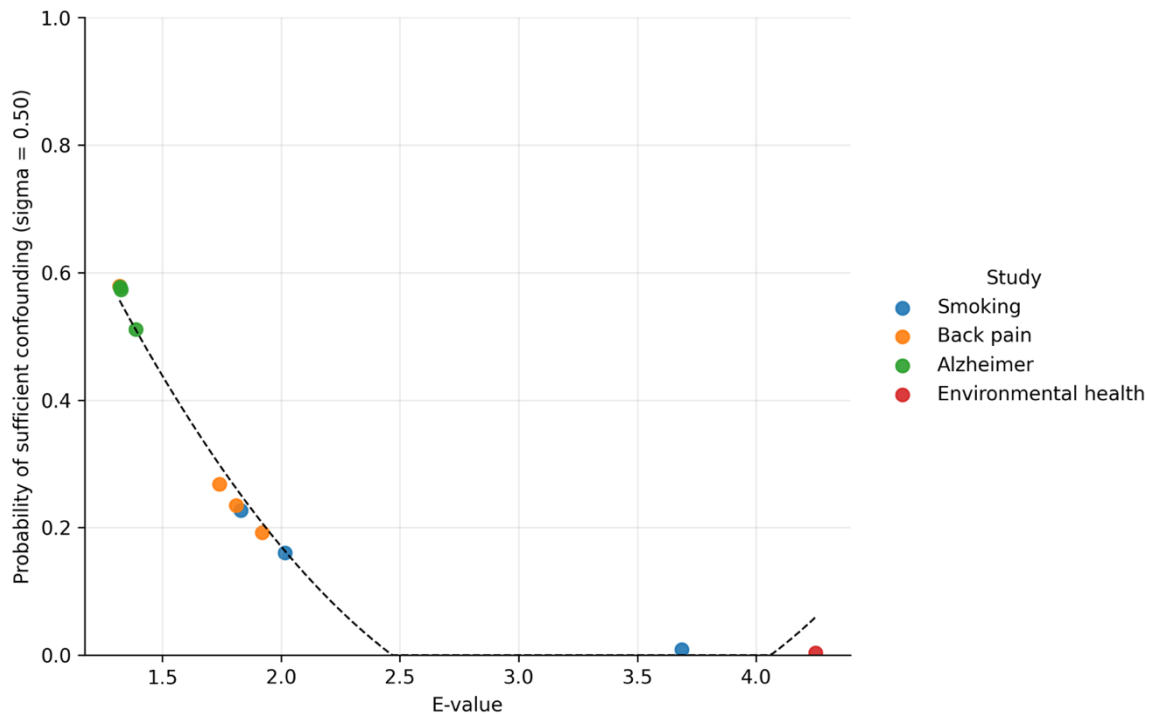

**Figure 1.** Relationship between the E-value and the posterior probability that unmeasured confounding is sufficiently strong to explain away the observed association, under a moderate prior ($\sigma = 0.5$). Each point represents a case study, with colors indicating different study domains. The dashed line represents a non-parametric smooth trend.

At the level of individual case studies, the results reveal a pronounced gradient of robustness. The environmental health study, characterized by the largest E-value ($E = 4.25$), yields an extremely low probability that unmeasured confounding could fully explain away the observed association ($P \approx 0.004$). Similarly, the strongest association in the smoking study ($E = 3.686$) corresponds to a low probability ($P \approx 0.03$), suggesting substantial robustness to unmeasured confounding.

In contrast, associations with smaller E-values show a markedly different behavior. The Alzheimer's disease case study, with E-values in the range 1.32–1.39, yields probabilities around 0.5, indicating that confounding of sufficient strength is not only possible but relatively plausible under the assumed prior. A similar pattern, though less extreme, is observed in the back pain study, where probabilities range from approximately 0.15 to 0.58, depending on the magnitude of the effect.

Intermediate results are observed within the smoking study when weaker exposure contrasts are considered. For example, maternal and household smoking yield probabilities around 0.16–0.20, indicating that while strong confounding is not highly likely, it cannot be ruled out with high confidence. These cases highlight the ability of the proposed framework to differentiate within a single study, rather than producing a uniform qualitative judgment.

A summary at the study level reinforces these patterns. The environmental health study emerges as highly robust, followed by the smoking study with moderate robustness. The back pain study occupies an intermediate position, while the Alzheimer's study appears substantially more sensitive to unmeasured confounding. This ranking is consistent with the ordering implied by the E-values, but provides a more interpretable probabilistic scale.

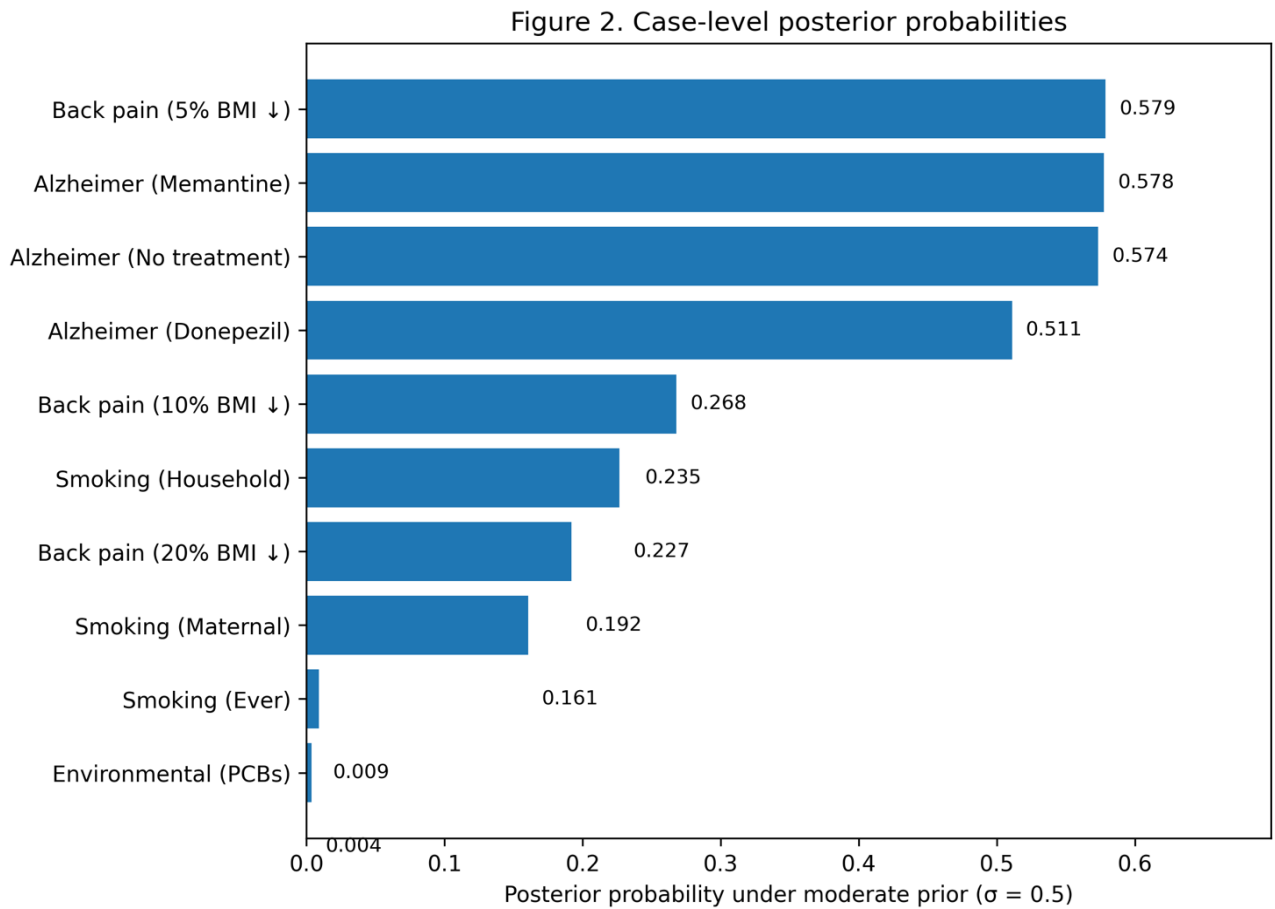

**Figure 2.** Case-level posterior probabilities that unmeasured confounding exceeds the E-value threshold under a moderate prior on confounding strength ( $\sigma = 0.5$ ). Cases are ordered by posterior probability, from highest to lowest. Shortened labels are used for readability, while preserving the original study domains and exposure contrasts. The figure highlights substantial heterogeneity both within and across domains, showing that some associations remain highly vulnerable to plausible confounding whereas others appear considerably more robust.

Figure 2 further illustrates the distribution of probabilities across individual cases, making explicit the heterogeneity both within and across studies. In particular, the contrast between high-E-value and low-E-value scenarios becomes visually evident, supporting the interpretation that the E-value operates as a boundary condition within a continuous probabilistic landscape.

Finally, sensitivity to prior assumptions was examined by varying the scale parameter governing the distribution of confounding strength. As shown in Figure 3, the qualitative conclusions remain stable across a wide range of priors for cases with large E-values, while they are more sensitive in cases with small E-values. This asymmetry reflects a key feature of the framework: strong observed effects remain robust under a broad class of assumptions, whereas weak effects are inherently more dependent on prior beliefs about confounding.

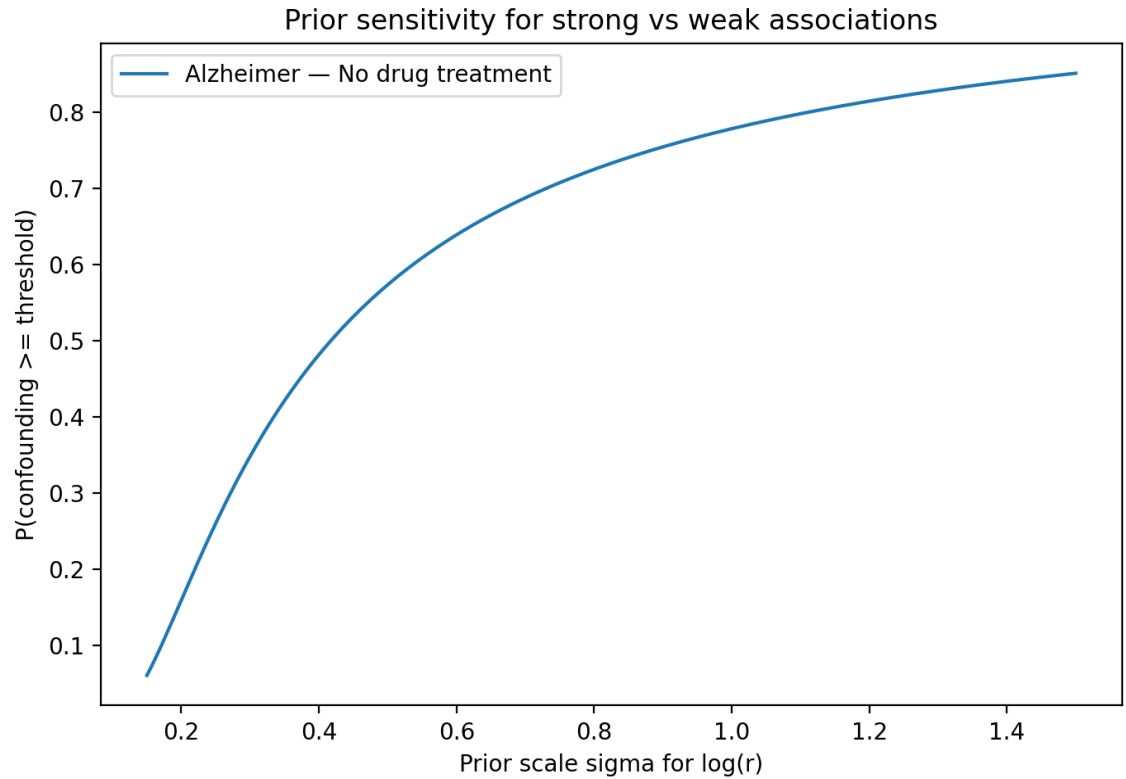

**Figure 3.** Sensitivity of posterior probabilities to prior assumptions on confounding strength. The figure compares selected case studies under different prior scales (σ = 0.25, 0.5, 1.0), illustrating that results for large E-values are stable, whereas small E-values are more sensitive to prior specification.

Taken together, these results demonstrate that the proposed Bayesian framework preserves the interpretability of the E-value while extending it into a fully probabilistic domain. Rather than classifying associations as robust or not based on heuristic thresholds, the method provides a direct quantification of how plausible it is that unmeasured confounding could explain away the observed effect.

Importantly, the relationship between E-values and posterior probabilities is not purely mechanical but depends on the assumed prior distribution of confounding strength, highlighting the role of prior knowledge in sensitivity analysis.

This highlights that the E-value alone does not provide an inferential measure of robustness, whereas the proposed probability explicitly quantifies the uncertainty associated with unmeasured confounding.

# Discussion

The present work proposes a conceptual and methodological shift in the way sensitivity analysis for unmeasured confounding is conducted. Rather than treating robustness as a threshold-based

property—defined by whether a hypothetical confounder could, in principle, explain away an observed association—we recast the problem in probabilistic terms. Specifically, we quantify the probability that confounding is sufficiently strong to exceed the E-value threshold, thereby transforming a deterministic diagnostic into a coherent inferential statement.

This shift addresses a well-recognized limitation of existing approaches. The Cornfield argument (Cornfield et al., 1959) and its modern implementation via the E-value (VanderWeele & Ding, 2017) provide clear and interpretable benchmarks for the strength of confounding required to invalidate an association. However, they do not answer the question that is often most relevant in practice: how plausible is it that such a level of confounding exists? In applied settings, this question is typically resolved through informal reasoning, relying on domain expertise without a formal probabilistic structure.

The proposed probability can be interpreted as a decision-relevant measure of evidential vulnerability, quantifying how likely it is that unmeasured confounding is sufficient to alter causal conclusions. By introducing a prior distribution over confounding strength, the present framework makes this assessment explicit and formally grounded. The resulting quantity, $P(\Gamma \geq \Gamma^*)$, provides a direct probabilistic statement about the robustness of the observed association. This interpretation aligns naturally with a Bayesian view of inference, in which uncertainty is expressed in terms of degrees of belief rather than long-run frequencies, thereby extending sensitivity analysis from a descriptive diagnostic to a coherent inferential tool.

The empirical results illustrate the practical implications of this reformulation. While the ordering of robustness across studies is broadly consistent with that implied by E-values, the probabilistic scale provides a more nuanced interpretation. In particular, the framework differentiates not only between studies, but also within studies, revealing heterogeneity that is obscured by single qualitative summaries. Moreover, the results highlight an important asymmetry: large E-values correspond to consistently low probabilities of sufficient confounding across a wide range of prior assumptions, whereas small E-values lead to conclusions that are more sensitive to prior specification. This behavior is not a limitation of the method, but rather a reflection of the intrinsic uncertainty associated with weak observed effects.

The proposed framework is closely related to previous efforts to probabilize sensitivity analysis. Probabilistic bias analysis (Greenland, 1996, 2005; Lash et al., 2009) and Bayesian models for unmeasured confounding (McCandless et al., 2007; Gustafson, 2010) already incorporate uncertainty in bias parameters. However, these approaches typically require detailed modeling assumptions and are often difficult to apply when only summary data are available. In contrast, the present method operates directly on quantities such as the E-value, making it applicable in settings where only aggregate information is accessible.

The approach also extends recent work by Costa (2025), which introduced a simulation-based perspective on the plausibility of confounding sufficient to explain away observed associations. While that framework provided an initial bridge between threshold-based reasoning and probabilistic interpretation, it relied on Monte Carlo estimation without an explicit generative model. The present work formalizes this intuition within a Bayesian framework, allowing prior assumptions to be clearly specified and systematically explored.

Several limitations should be acknowledged. First, the framework relies on the specification of a prior distribution over confounding strength, which inevitably introduces subjectivity. However, this subjectivity is not a drawback per se; rather, it reflects the unavoidable role of background knowledge in assessing the plausibility of unmeasured confounding. Importantly, the sensitivity analysis with respect to the prior allows this dependence to be made transparent.

Second, the model adopts a simplified representation of confounding through a single parameter. While this abstraction facilitates interpretation and computation, real-world confounding structures may be more complex, involving multiple variables and interactions. Extending the framework to richer representations of confounding is a natural direction for future work.

Third, the empirical analysis is based on a limited set of case studies. Although these examples span multiple domains and illustrate the key features of the method, a broader evaluation across a larger corpus of observational studies would be valuable to assess generalizability.

More broadly, this framework aligns sensitivity analysis with Bayesian principles of inference under uncertainty, replacing threshold-based reasoning with probabilistic statements that can directly inform scientific and clinical judgment.

Despite these limitations, the proposed framework offers a conceptually coherent and practically useful extension of existing methods. It preserves the interpretability of the E-value while addressing its central limitation: the absence of a probabilistic interpretation.

# Conclusions

This study introduces a Bayesian reformulation of sensitivity analysis for unmeasured confounding, in which the strength of confounding is treated as a random variable rather than an unknown constant. By linking the E-value to a probabilistic measure $P(\Gamma \geq \Gamma^*)$, the approach provides a direct quantification of how plausible it is that unmeasured confounding could explain away an observed association.

This perspective transforms sensitivity analysis from a threshold-based diagnostic into a fully inferential procedure. Instead of asking whether a confounder of sufficient strength could exist, the proposed method allows researchers to assess how likely such a confounder is, given explicit assumptions. In doing so, it bridges a critical gap between methodological tools and the interpretative questions that arise in applied research.

More broadly, the framework contributes to a shift in the logic of causal inference from binary judgments to probabilistic reasoning. By making assumptions explicit and quantifying their implications, it aligns sensitivity analysis with a coherent theory of inference under uncertainty. This alignment is particularly important in observational research, where unmeasured confounding cannot be eliminated, but only reasoned about.

Future work may extend this approach in several directions, including the development of hierarchical models for confounding, integration with individual-level Bayesian analyses, and applications to large-scale evidence synthesis. Ultimately, the goal is not to eliminate uncertainty, but to represent it in a form that supports transparent and rational scientific judgment.